\documentclass[epj]{svjour}
\usepackage{amsmath}
\usepackage{amssymb}
\usepackage{graphicx}
\usepackage[numbers,sort&compress]{natbib}
\usepackage{units}

\begin{document}

\title{Interaction between concentric Tubes in DWCNTs}

\author{R. Pfeiffer\inst{1}\thanks{\email{rpfei@ap.univie.ac.at}} \and Ch.
  Kramberger\inst{1,2} \and F.  Simon\inst{1} \and H. Kuzmany\inst{1} \and V.
  N. Popov\inst{3} \and H.  Kataura\inst{4}}

\institute{Institut f\"ur Materialphysik, Universit\"at Wien, Strudlhofgasse
  4, A-1090 Wien, Austria \and Leibniz Institute of Solid State and Materials
  Research Dresden, Helmholtzstra\ss{}e 20, D-01069 Dresden, Germany \and
  Faculty of Physics, University of Sofia, Sofia, Bulgaria \and Tokyo
  Metropolitan University, 1-1 Minami-Ohsawa, Hachioji, Tokyo 192-0397, Japan}

\date{\today}

\abstract{A detailed investigation of the Raman response of the inner tube
  radial breathing modes (RBMs) in double-wall carbon nanotubes is reported.
  It revealed that the number of observed RBMs is two to three times larger
  than the number of possible tubes in the studied frequency range. This
  unexpected increase in Raman lines is attributed to a splitting of the inner
  tube response. It is shown to originate from the possibility that one type
  of inner tube may form in different types of outer tubes and the fact that
  the inner tube RBM frequency depends on the diameter of the enclosing tube.
  Finally, a comparison of the inner tube RBMs and the RBMs of tubes in
  bundles gave clear evidence that the interaction in a bundle is stronger
  than the interaction between inner and outer tubes.
   }

\maketitle

\section{Introduction}

Single-wall carbon nanotubes (SWCNTs)
\citep{Iijima:Nature363:603:(1993),Thess:Science273:483:(1996)} have
attracted a lot of scientific interest over the last decade due to
their unique structural and electronic properties.  In
\citeyear{Smith:Nature396:323:(1998)},
\citet{Smith:Nature396:323:(1998)} discovered that fullerenes can be
encapsulated in SWCNTs, forming so-called peapods
\citep{Smith:ChemPhysLett321:169:(2000),Kataura:SyntheticMet121:1195,%
Simon:ChemPhysLett383:362:(2004)}.  By annealing such peapods at high
temperatures in a dynamic vacuum the enclosed C$_{60}$ peas transform
into SWCNTs within the outer tubes, thus producing double-wall carbon
nanotubes (DWCNTs) \citep{Bandow:ChemPhysLett337:48:(2001),%
Bandow:PhysRevB66:075416:(2002)}. The growth process of the inner
tubes is a new route for the formation of SWCNTs under shielded
conditions in the absence of any additional catalyst.
  
A detailed Raman study of the radial breathing modes (RBMs) of the inner tubes
revealed that these modes have intrinsic linewidths down to
$\unit[0.4]{cm^{-1}}$ \citep{Pfeiffer:PhysRevLett90:225501:(2003)}. These
small linewidths indicate long phonon lifetimes and therefore highly defect
free inner tubes. Thus, they are a direct experimental evidence for a nano
clean room reactor on the inside of SWCNTs.

Peapod grown DWCNTs are also interesting from another point of view.  Usually,
the peapods are produced from standard SWCNTs with diameters around
$\unit[1.4]{nm}$. Taking into account the van der Waals interaction between
the walls this means that the diameters of the inner shell tubes are around
$\unit[0.7]{nm}$. For these thin tubes the possible diameters can no longer be
assumed to form a quasi-continuum. Additionally, due to the inverse relation
between RBM frequency and diameter, the spectral distance between the RBMs of
different inner tube types is (with few exceptions) much larger than between
different outer tube RBMs. This opens the possibility to study the properties
of individual SWCNTs in a bulk sample. A first application of this was the
assignment of the chiral vectors to all inner tubes
\citep{Kramberger:PhysRevB68:235404:(2003)}.

Studying the Raman response of the inner tube RBMs in high resolution at
$\unit[20]{K}$ revealed that the number of observed lines is about two to
three times larger than the number of geometrically allowed inner tubes. In
the following, we show that this unexpected increase of Raman lines can be
attributed to a splitting of the response from the inner tubes. This splitting
originates from the possibility that one inner tube type may form in more than
one outer tube type (with different diameters) and the fact that the RBM
frequency of the inner tube depends on the diameter of the enclosing parent
tube. A quantitative explanation for the splitting was obtained by calculating
the RBM frequencies of the inner tubes as a function of the outer tube
diameter within a continuum model. Using two different scenarios regarding the
possible inner--outer shell pairs, we compared the splitting obtained from the
simulation with the experimentally obtained splitting. From this we conclude
that not only the best fitting inner tubes are formed.

\section{Experimental}

The starting material for the DWCNTs were C$_{60}$ peapods (in the form of
bucky paper) produced with a previously described method
\cite{Kataura:SyntheticMet121:1195}. This DWCNT sample is the one discussed in
Ref.~\citealp{Pfeiffer:PhysRevLett90:225501:(2003)}. The mean diameter of the
assumed Gaussian diameter distribution of the outer tubes was
$\unit[1.39(2)]{nm}$ with a variance of $\unit[0.1]{nm}$, as determined from
the RBM Raman response \citep{Kuzmany:EurPhysJB22:307:(2001)}. The filling of
the tubes large enough for C$_{60}$ to enter was close to $\unit[100]{\%}$ as
evaluated from a Raman \citep{PfeifferKB2002,Kuzmany:ApplPhysA76:449(2003)}
and EELS analysis \citep{Liu:PhysRevB65:045419:(2002)}. These peapods were
slowly heated up to $\unit[1280]{^\circ C}$ in a dynamic vacuum, annealed for
$\unit[2]{h}$, and were then slowly cooled down to room temperature.

The Raman spectra were measured with a Dilor xy triple spectrometer using
various lines of an Ar/Kr laser, a He/Ne laser and a Ti:sapphire laser. The
spectra were recorded at $90$ and $\unit[20]{K}$ in normal (NR) and high
resolution (HR) mode, respectively
($\Delta\bar{\nu}_{\text{NR}}=\unit[1.3]{cm^{-1}}$ and
$\Delta\bar{\nu}_{\text{HR}}=\unit[0.4]{cm^{-1}}$ in the red). In these
measurements the samples were glued on a copper cold finger with silver paste.

\section{Theory}

The RBM frequencies of a given DWCNT were calculated using a continuum
model (model $2$ in Ref.~\citealp{Popov:PhysRevB65:235415:(2002)}). In
this model, the DWCNT is represented by two nested hollow cylinders
with diameters $d_{\text{i}}$ and $d_{\text{o}}$ of the inner and
outer shells, respectively.  The interaction between two points at a
distance $r$ on different shells was described by a Lennard--Jones
(LJ) potential
\begin{equation}
V(r)=4\epsilon[(\sigma/r)^{12}-(\sigma/r)^6]\,,
\end{equation}
where $\epsilon=\unit[2.964]{meV}$ and $\sigma=\unit[0.3407]{nm}$
\citep{Lu:PhysRevLett68:1551:(1992)}.

The total interaction energy $\Phi$ between the two shells was
calculated by a numerical integration over unit length of the
shells. For this the two shells were approximated by quadratic
meshes. The mass of the carbon atoms was equally distributed over the
meshes and located in their centers. The interaction energy was then
calculated by summing the LJ potential over all mesh centers. The size
of the meshes was reduced until the interaction energy had converged.

The interaction part of the dynamical matrix was then obtained by a
numerical differentiation of $\Phi$ with respect to $d_{\text{i}}$ and
$d_{\text{o}}$. It was also assumed that the separate shells are
elastic and are characterized with force constants $k_{\text{i}}$ and
$k_{\text{o}}$ for the radial breathing motion.  The $k$'s were
determined from the RBM frequency $\bar{\nu}_{\text{RBM}}$ of a tube
which is related to the tube diameter $d$ by
\citep{Jishi:ChemPhysLett209:77:(1993),%
Kuerti:PhysRevB58:R8869:(1998),Henrard:PhysRevB60:R8521:(1999)}
\begin{equation}
\label{Eq:RBM}
\bar{\nu}_{\text{RBM}}=C_1/d+C_2\,,
\end{equation}
where $C_1$ and $C_2$ are constants. The role of $C_2$ is to account for all
frequency shifts due to the interaction with the environment. Since this
interaction was explicitly modeled in our simulations, $C_2$ was set zero in
the calculations. The shell diameters were taken from a DFT study
\citep{Kramberger:PhysRevB68:235404:(2003)}, giving
\begin{equation}
\label{Equ:dDFT}
d=\left(\frac{1}{d_{\text{G}}}-\frac{0.0050}{d_{\text{G}}^2}-
\frac{0.0013}{d_{\text{G}}^4}\right)^{-1}\,,
\end{equation}
where
\begin{equation}
d_{\text{G}}=\frac{\sqrt{3}\,a_{\text{CC}}}{\pi}\sqrt{m^2+mn+n^2}
\end{equation}
is the graphene folding diameter, $a_{\text{CC}}=\unit[0.141]{nm}$ is the
C$-$C distance in graphene and $(m,n)$ is the chiral vector of the tube.
Especially for small diameter tubes ($d\lessapprox\unit[1]{nm}$) the DFT
derived diameter instead of the graphene folding diameter has to be used.

The dynamical matrix $D$ of the DWCNT can then be written in the form
\begin{equation}
D=-\frac{1}{\sqrt{m_{\text{i}}m_{\text{o}}}}
\begin{pmatrix}
k_{\text{i}}+\partial^2\Phi/\partial d_{\text{i}}^2 & 
\partial^2\Phi/\partial d_{\text{i}}\partial d_{\text{o}}\\
\partial^2\Phi/\partial d_{\text{o}}\partial d_{\text{i}} &
k_{\text{o}}+\partial^2\Phi/\partial d_{\text{o}}^2\\
\end{pmatrix}
\,,
\end{equation}
where $m_{\text{i}}$ and $m_{\text{o}}$ are the masses (per unit length) of
the inner and outer shells, respectively. We note in passing that this form of
$D$ differs from that in Ref.~\citealp{Dobardzic:PhysStatSolB237:R7:(2003)},
where the partial derivatives have been assumed equal. The RBM frequencies are
finally obtained as solutions of the vibrational eigenvalue problem for the
DWCNT.

The results of the relaxation of the DWCNTs with respect to the displacement
of the shell axis from the coaxial position show that the two shells stay
coaxial for $\Delta d=d_{\text{o}}-d_{\text{i}}<\unit[0.78]{nm}$ and that they
are no longer coaxial for $\Delta d>\unit[0.78]{nm}$. For coaxial shells, the
relaxation of the DWCNT with respect to $d_{\text{i}}$ (for fixed
$d_{\text{o}}$) or vise versa yielded an equilibrium value for $\Delta d$ of
about $\unit[0.68]{nm}$.

For the following analysis, we calculated the RBM frequencies for all
inner--outer tube pairs with inner tubes between $(5,3)$
($d_{\text{i}}=\unit[0.55]{nm})$ and $(14,0)$ ($d_{\text{i}}=\unit[1.09]{nm}$)
and possible diameter differences in the range $0.66$--$\unit[0.74]{nm}$.

\section{Experimental Results}

\begin{figure}
\includegraphics[width=\linewidth,clip]{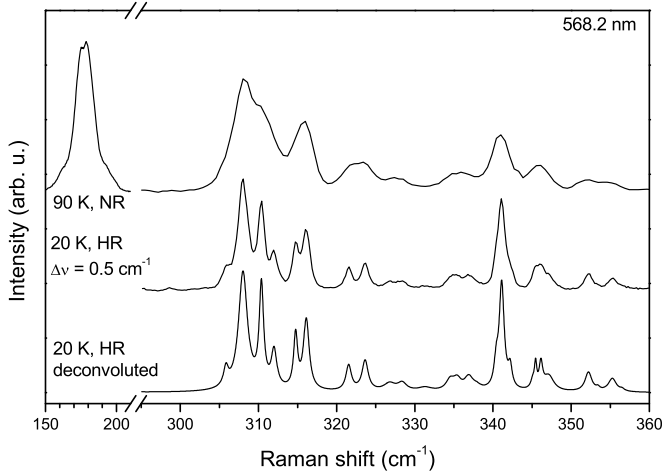}
\caption{RBM Raman response of the inner tubes at $\unit[90]{K}$ and
  normal resolution, $\unit[20]{K}$ and high resolution, and
  Lorentzian components of the $\unit[20]{K}$ and high resolution
  spectrum (top to bottom).}
\label{Fig:G568}
\end{figure}

Figure~\ref{Fig:G568} depicts the Raman response of the inner tube
RBMs. While the NR spectrum shows only very broad lines, the HR
spectrum reveals much richer details with rather small
linewidths. This becomes even more obvious when the as measured
spectrum was fitted with Voigtian lines (Lorentzians convoluted with a
Gaussian spectrometer response) where the spectrometer response was
taken from a fit to the elastically scattered light. The resulting
Lorentzian components are equivalent to a deconvoluted spectrum
\citep{Pfeiffer:PhysRevLett90:225501:(2003)}.

Using the diameter--frequency relationship from Eq.~\eqref{Eq:RBM} it
is possible to determine the number of inner tubes for a given
frequency range. There is some ongoing discussion on the value of
$C_1$, where reported numbers range from $224$ to
$\unit[250]{cm^{-1}\,nm}$ \citep{Jorio:PhysRevLett86:1118:(2001),%
Bachilo:Science298:2361:(2002),Kramberger:PhysRevB68:235404:(2003)}. In
the following, we use the value of $\unit[233]{cm^{-1}\,nm}$ as
derived from a systematic analysis of the inner shell tubes in DWCNTs
\citep{Kramberger:PhysRevB68:235404:(2003)}.

Regardless of the exact value of $C_1$, between $300$ and
$\unit[350]{cm^{-1}}$ one should only see the response of eight different
tubes. By looking at the HR spectrum in figure~\ref{Fig:G568}, one can easily
identify about $18$ peaks and shoulders in this range. Hence, one can observe
about two to three times more RBMs as there are geometrically allowed tubes.

\begin{figure}
\includegraphics[width=\linewidth,clip]{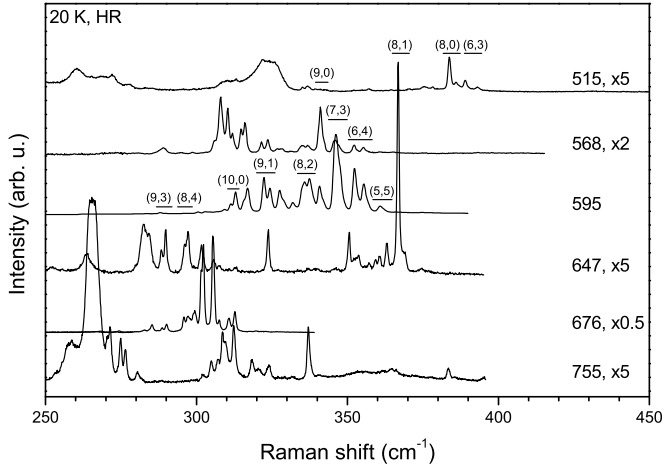}
\caption{High resolution Raman response of the inner tube RBMs for
  several excitation wavelengths at $\unit[20]{K}$. Selected chiralities after
  Ref.~\citealp{Kramberger:PhysRevB68:235404:(2003)}. Split widths are
  indicated by lines under the chirality vectors.}
\label{Fig:RBM}
\end{figure}

Figure~\ref{Fig:RBM} depicts selected high resolution Raman spectra of the
inner RBMs. Using the refined frequency--dia\-meter relation from
Ref.~\citealp{Kramberger:PhysRevB68:235404:(2003)}, one should find the RBMs
of $28$ distinct inner tubes between $270$ and $\unit[400]{cm^{-1}}$. Again
the observed number of lines in this region is about three times larger. For
some chiralities the split widths are indicated by lines under the folding
vector components.

\begin{figure}
\centering
\includegraphics[width=\linewidth]{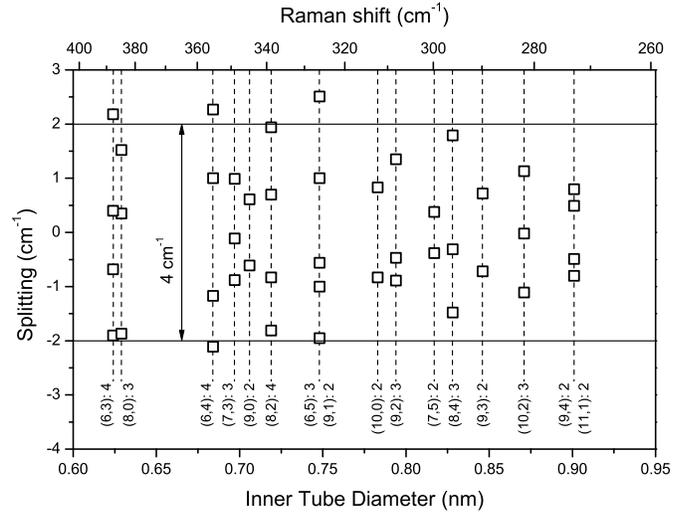}
\caption{Number of split components and width of splitting for
  selected individual inner tubes from the spectra in figure~\ref{Fig:RBM}
  after subtracting the mean value of the RBM frequencies. The dashed lines
  connect split components assigned to the same inner tube. Additionally, the
  tube chiralities and the number of split components is included.}
\label{Fig:SplitExp}
\end{figure}

In order to determine the number of the split components and the width of the
splitting, we fitted the spectra from figure~\ref{Fig:RBM} with a number of
Voigtian lines. Using the chirality assignment of
Ref.~\citealp{Kramberger:PhysRevB68:235404:(2003)}, sets of RBM frequencies
were assigned to selected inner tube types. The tubes were selected such that
the assignment was unambiguous. If the frequency sets of two inner tubes
overlapped the tubes were not considered in the following analysis.

For the selected inner tubes, we subtracted the mean value from the assigned
frequencies and plotted the split components vs.\@ inner tube diameter in
figure~\ref{Fig:SplitExp}. It shows that the number of split components ranges
between two and four with no obvious influence of the tube chirality.
Additionally, the width of the splitting is about $\unit[4]{cm^{-1}}$. Since
not all inner tubes were considered in the analysis, this splitting value is a
lower limit.

\section{Simulation Results}

\begin{figure}
\includegraphics[width=\linewidth,clip]{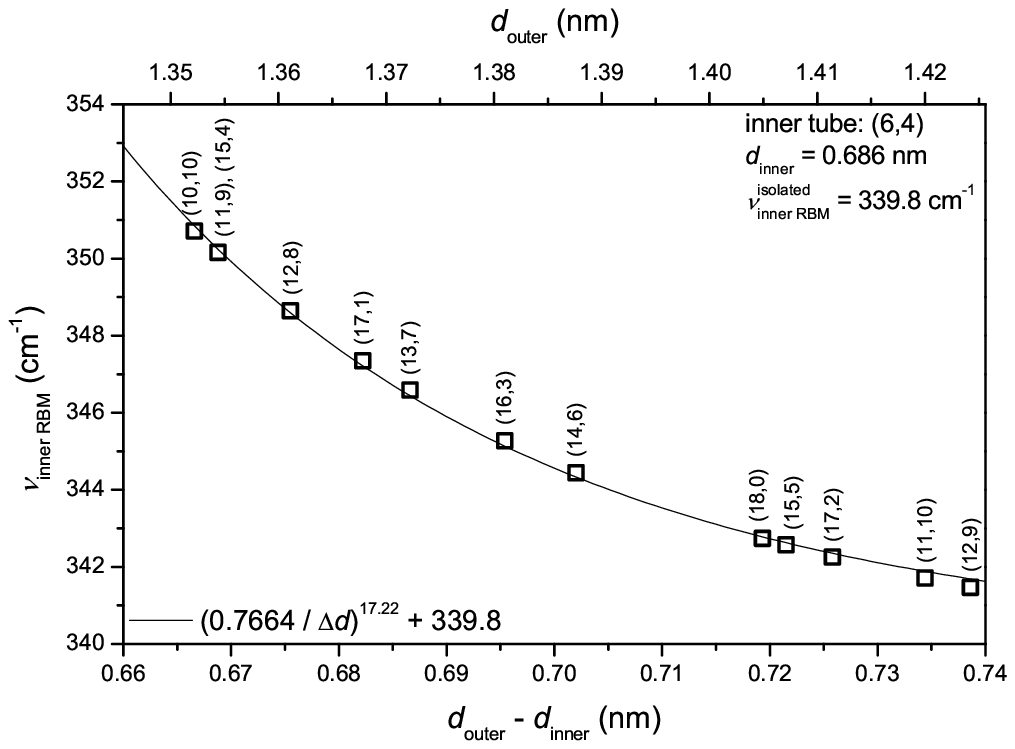}
\newline
\includegraphics[width=\linewidth,clip]{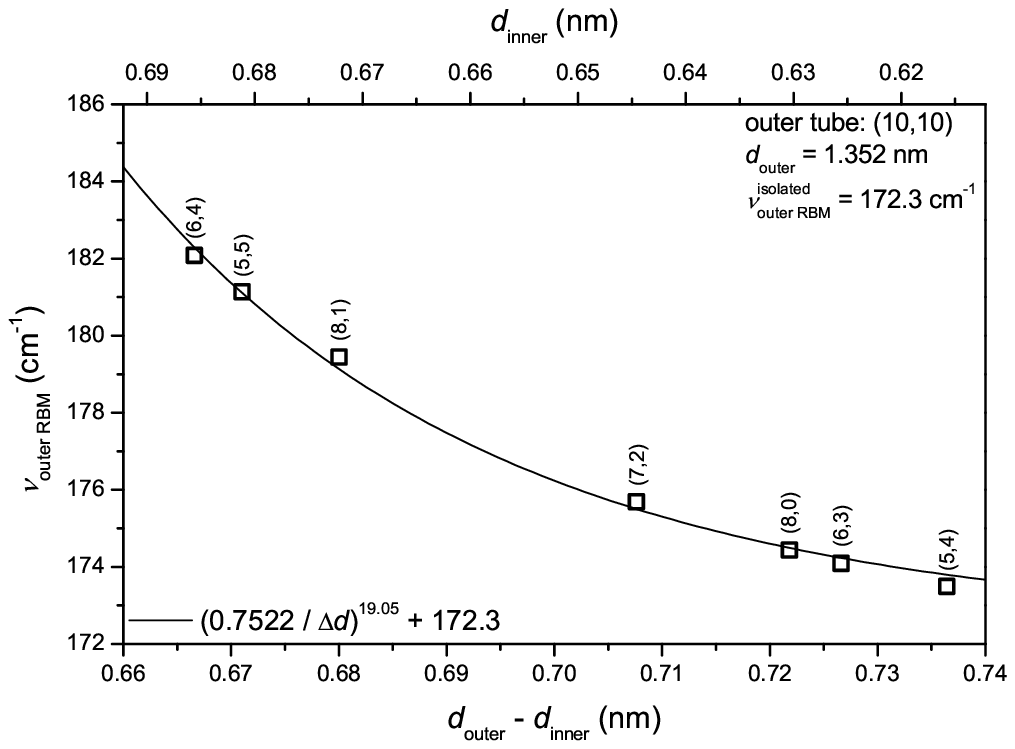}
\caption{Top: Calculated RBM frequency of a $(6,4)$ inner tube as a
  function of the encapsulating tube diameter $d_{\text{o}}$. Bottom:
  Calculated RBM frequency of a $(10,10)$ outer tube as a function of the
  encapsulated tube diameter $d_{\text{i}}$. The diameters were calculated
  from Eq.~\eqref{Equ:dDFT}. The lines are fits with the given parameters as
  indicated in the lower left corners.}
\label{Fig:ITOT}
\end{figure}

Figure~\ref{Fig:ITOT} (top) demonstrates the dependence of the calculated RBM
frequency of a $(6,4)$ inner tube on the diameter of the outer tube. The
isolated $(6,4)$ tube has a diameter of $d_{\text{i}}=\unit[0.686]{nm}$ and a
frequency of $\bar{\nu}_{\text{inner
    RBM}}^{\text{isolated}}=\unit[339.8]{cm^{-1}}$ (excluding $C_2$)
\citep{Kramberger:PhysRevB68:235404:(2003)}. Due to the interaction between
the two shells, the inner RBM frequency increases by up to
$\unit[12]{cm^{-1}}$ with decreasing outer tube diameter. In the diameter
difference range studied, the RBM frequencies can be fitted with $(a/\Delta
d)^b+\bar{\nu}_{\text{RBM}}^{\text{isolated}}$ (see figure~\ref{Fig:ITOT},
lower left corners).

Similarly, figure~\ref{Fig:ITOT} (bottom) depicts the RBM frequency of a
$(10,10)$ outer tube as a function of the inner tube diameter. Again, the
larger the inner tube diameter the larger the shift of the outer tube RBM.
Indeed, slight upshifts of a few $\unit{cm^{-1}}$ of the outer tube RBMs of
DWCNTs were observed when directly compared with the empty reference SWCNTs.
However, due to the quasi-continuous diameter distribution of the outer tubes,
exact measurements are difficult.

\begin{figure}
\includegraphics[width=\linewidth,clip]{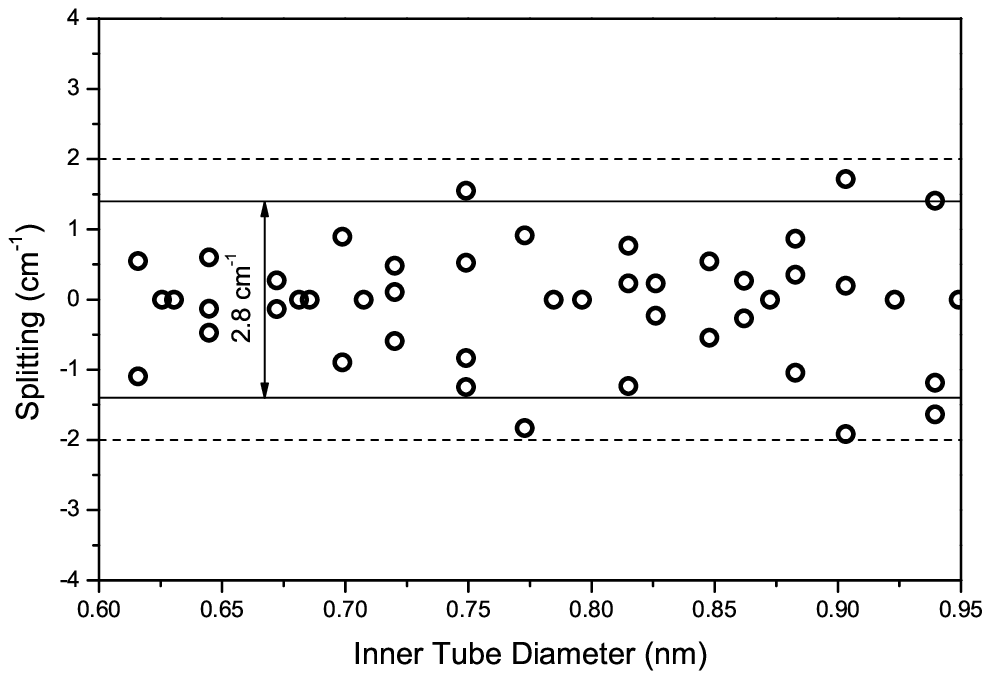}
\newline
\includegraphics[width=\linewidth,clip]{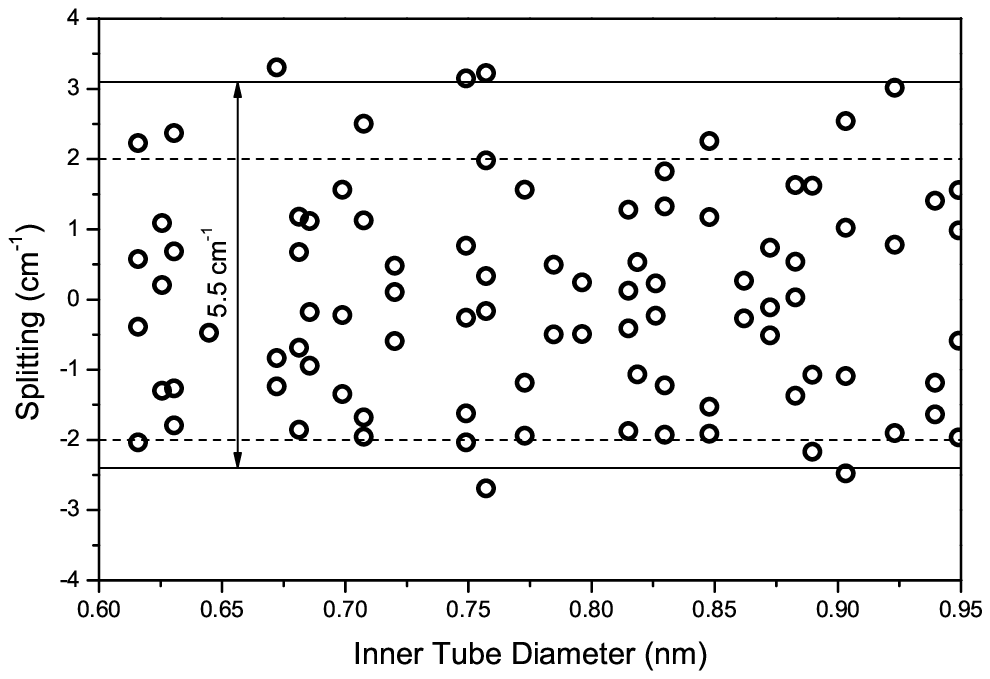}
\caption{Number of split components and width of splitting for
  individual inner tubes from the theory after subtracting the mean value of
  the calculated RBM frequencies. Top: Only the best fitting inner tubes are
  formed. Bottom: The best and second best fitting inner tubes are formed. The
  dashed lines mark the width of splitting obtained from the experiment.}
\label{Fig:SplitTheo}
\end{figure}

In a first step, we assumed that in all outer tubes in our sample only
the best fitting inner tubes are formed. Best fitting means that for
every outer tube the inner tube was selected such that
$|d_{\text{outer}}-d_{\text{inner}}-\unit[0.68]{nm}|\to\min$.  The
splitting calculated for this assumption is depicted in
figure~\ref{Fig:SplitTheo} (top). The number of split components and
the width of the splitting are smaller than the experimentally
observed values.  Therefore, in a second step, we assumed that also
the second best fitting inner tubes may form. As
figure~\ref{Fig:SplitTheo} (bottom) shows, this assumption results in
a splitting of $\unit[5.5]{cm^{-1}}$ which is larger than
observed. This suggests that also second best fitting inner tubes form
but only in cases where the energy balance is not too bad.

\begin{figure}
\includegraphics[width=\linewidth,clip]{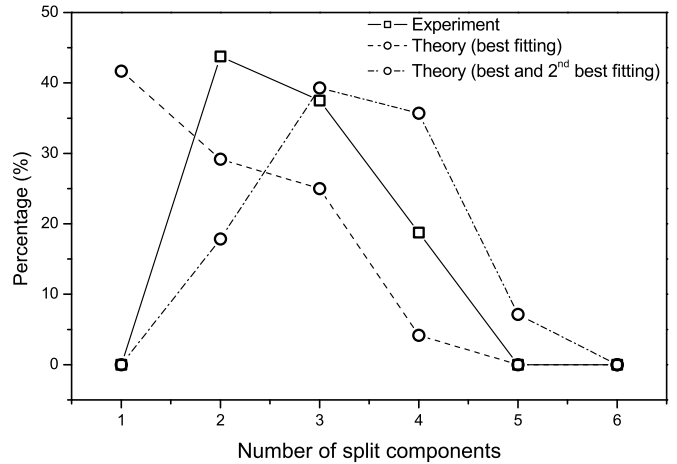}
\caption{Percentages of the number of split components from the
  experiment and the two theoretical models.}
\label{Fig:relHfgk}
\end{figure}

Figure~\ref{Fig:relHfgk} compares the percentages of the number of
inner tube split components obtained from the experiment with the two
theoretical models discussed above. If only best fitting inner tubes
form, mainly $1$, $2$ and $3$ split components should be observed. If
best and second best fitting inner tubes form, mainly $3$ and $4$
split components are to be expected. The experimental curve lies
between the two model curves peaking around $2$--$3$ split components.

\section{Discussion}

The continuum model we are using was parameterized for graphite
\citep{Lu:PhysRevLett68:1551:(1992)} and subsequently used for
simulations of multi-shell fullerenes
\citep{Lu:PhysRevB49:11421:(1994)} and multi-wall carbon nano\-tubes
\citep{Popov:PhysRevB65:235415:(2002)}. As mentioned above, it
resulted in an equilibrium diameter difference of about
$\unit[0.68]{nm}$.  \citet{Abe:PhysRevB68:041405(R):(2003)} reported
X-ray measurements that suggest a more likely diameter difference of
$\unit[0.72]{nm}$. Even though a recent Raman analysis supports this
value \citep{Simon:cond-mat0403179}, we think that our main results
are not affected by this difference. For larger equilibrium diameters
the interaction will be weaker in general and therefore more likely
reduce the calculated splitting.

In the simulations we were using, only the diameter dependence of the RBM
frequency shift can be evaluated but not the influence of differing
chiralities of the inner and outer tubes in one DWCNT. This simplification is
justified by the experimental results that show no obvious dependence of the
splitting on the inner tube chirality. Therefore, we conclude that the
chirality plays only a minor role. On the other hand, an influence of the
chirality on the interaction energy may not be completely negligible. It can
be the reason for the observed best and second best fitting tube grown inside
the outer tube. Accepting this as the reason for the extended splitting has
important consequences for the growth dynamics.

Always these tubes will grow which are energetically most favorable,
at the high transformation temperatures. This holds even if nucleation
of inner tube growth starts at different positions in the outer
tube. Interesting configurations occur for the growth of a chiral tube
inside an achiral tube.  There, the two possible stereo-isomers are
equal in energy but cannot match.  The proven low concentration of
defects on the inner tubes evidences that always one of the
equienergetic tubes will win.

Our calculations have shown that the upshift of the inner tube RBMs can be as
high as $\unit[12]{cm^{-1}}$. Such an upshift of the RBM frequency does not
only occur for inner tubes in DWCNTs but also for SWCNTs in bundles
\citep{Henrard:PhysRevB60:R8521:(1999),Henrard:PhysRevB64:205403:(2001)}.  In
both cases this shift is due to the van der Waals interaction with the
surrounding tubes. However, in the case of the concentric tubes the
interaction is between a tube outside and a tube inside whereas in the case of
the bundles it is between two tube outsides. Therefore, it was interesting to
compare the upshift of encaged and bundled tubes. For this comparison we used
a typical HiPco sample with a broad diameter distribution such that similar
tube diameters could be compared.

\begin{figure}
\includegraphics[width=\linewidth,clip]{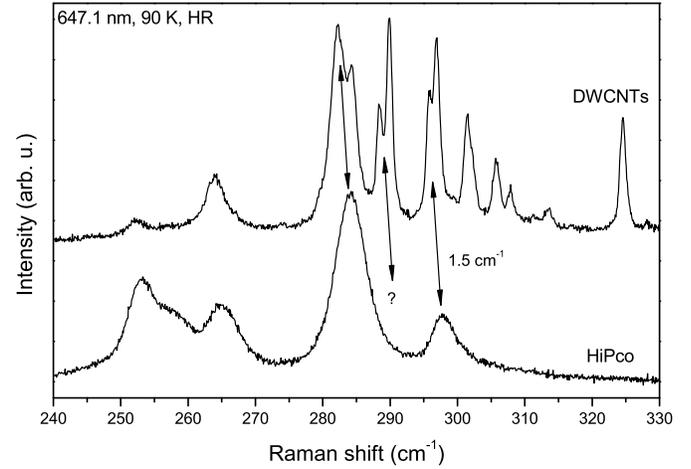}
\caption{RBM Raman response of small diameter tubes as inner tubes of
  DWCNTs (top) and in a typical HiPco sample (bottom). The HiPco RBMs are
  $\unit[1.5]{cm^{-1}}$ more upshifted than the inner tube RBMs. Additionally,
  the resonances of some bundled tubes are different to the corresponding
  encaged tubes.}
\label{Fig:HiPcoDWCNT}
\end{figure}

Figure~\ref{Fig:HiPcoDWCNT} depicts the RBM Raman response of the
inner tubes and the HiPco tubes. The two samples were measured
simultaneously in the same experiment. This means, the spectra were
recorded without changing the spectrometer position which guarantees a
high precision of the measured frequency differences. Several
observations can be made in this figure.

First, compared with the inner tube RBMs the HiPco RBMs are broader
and do not get narrower in HR mode. This is an important detail of the
results, since it demonstrates that the narrow linewidth is not a mere
consequence of the small tube diameter. It rather underlines the high
quality of the inner tubes. Additionally, the HiPco RBMs may be
broadened due to the inhomogeneous environment within the bundles
\citep{Henrard:PhysRevB64:205403:(2001)}.

Second, the HiPco RBMs are about $\unit[1.5]{cm^{-1}}$ more upshifted
than the inner tube RBMs. The larger upshift of the HiPco sample
suggests that the interaction between the tubes in the bundles is
stronger than between the inner and outer shell of a DWCNT. This
experimental evidence is rather surprising since the interaction area
for one tube in a bundle is certainly smaller than for inner--outer
tube pairs. Thus, the reason for the upshift is not yet fully
understood. It may be related to a larger equilibrium diameter
difference in the concentric tubes as compared to the distance between
tubes in the bundles. This increased distance would reduce the
interaction between inner and outer tubes.

Third, some RBMs are missing in the HiPco spectrum. Since the RBM
response of even thinner tubes than the missing species is present in
the HiPco spectrum the observed absence cannot be explained by the
diameter distribution of the HiPco material. In fact, the missing
lines show up for other excitation energies
\citep{Simon:cond-mat0404110}. This means, the resonances of the
encaged tubes are slightly different to the resonances in the bundled
tubes.

\section{Summary}

We have shown that the RBMs of the inner tubes of DWCNTs are split into
several components and provided a quantitative understanding from
calculations. The splitting is attributed to the interaction between inner and
outer tubes that causes a change of the inner tube RBM frequency. Since it is
possible that one type of inner tube forms in several types of outer tubes
(with slightly different diameters) every inner tube gives rise to more than
one RBM in the Raman spectrum. We have further demonstrated that not only the
best fitting structure is established. Finally, we have compared the RBM Raman
response of inner tubes with that of similar diameter tubes in bundles.
Surprisingly, the bundled tube RBMs are more upshifted than the encaged tube
RBMs.

\begin{acknowledgement}
  The authors acknowledge financial support from the FWF in Austria, project
  P14893, and from the EU Marie-Curie project MEIF-CT-2003-501099 (PATONN).
  VNP was supported by a fellowship from the Federal Science Policy Office for
  promoting the S\&T cooperation with Central and Eastern Europe, by a NATO
  CLG, and from the EU Marie-Curie project MEIF-CT-2003-501080.
\end{acknowledgement}

\bibliographystyle{apsrev}
\bibliography{Splitting}

\end{document}